\documentclass[12pt]{article} \usepackage{a4wide} \usepackage{amssymb}
\usepackage{graphicx}
\begin{document}
{\renewcommand{\thefootnote}{\fnsymbol{footnote}}
\begin{center}
{\LARGE  Canonical description of quantum dynamics}\\
\vspace{1.5em}
Martin Bojowald\footnote{e-mail address: {\tt bojowald@psu.edu}}
\\
\vspace{0.5em}
Institute for Gravitation and the Cosmos,\\
The Pennsylvania State
University,\\
104 Davey Lab, University Park, PA 16802, USA\\
\vspace{1.5em}
\end{center}
}

\setcounter{footnote}{0}

\begin{abstract}
  Some of the important non-classical aspects of quantum mechanics can be
  described in more intuitive terms if they are reformulated in a geometrical
  picture based on an extension of the classical phase space. This
  contribution presents various phase-space properties of moments describing a
  quantum state and its dynamics. An example of a geometrical reformulation of
  a non-classical quantum effect is given by an equivalence between conditions
  imposed by uncertainty relations and centrifugal barriers, respectively.
\end{abstract}
 
\section{Introduction}

Although classical and quantum mechanics are usually presented based on
different mathematical methods and with a marked contrast in the underlying
physical concepts, there are formulations that bring the two theories closer
to each other. For instance, the Koopman wave function \cite{Koopman} can be
used to make classical mechanics look quantum.  Effective theories reverse the
direction and make quantum mechanics look classical.

Given the richer nature of quantum mechanics with more degrees of freedom of a
state than in classical mechanics, there are different ways in which this
``classicalization'' can be achieved. A method of importance in elementary
particle physics is the low-energy effective action \cite{EffAcQM} which
determines the dynamics of a quantum system around the ground state, and the
related Coleman--Weinberg potential \cite{ColemanWeinberg}. Such low-energy
methods are useful in situations in which only a few particle degrees of
freedom are excited, as in basic scattering processes of interest in particle
physics. Methods that do not rely on low-energy regimes are geometric ones or
phase-space descriptions of quantum mechanics, going back for instance to the
symplectic formulations of \cite{RealImQuantum,GeomQuantMech,ClassQuantMech}. These
approaches make use of the fact that the imaginary part of an inner product is
antisymmetric and non-degenerate, and therefore can be interpreted as a
symplectic form on the space of states.

Here, we use a geometrical picture of quantum dynamics derived from a
suitable parameterization of states.  In some ways, it is related to
symplectic phase-space descriptions, but it includes a suitable generalization
that makes it possible to formulate effective theories in which certain
degrees of freedom contained in a full quantum state are suppressed. (The
parameterization of states to be introduced determines which degrees of
freedom may be suppressed.) Most of these truncations cannot be described by
symplectic methods because removing some degrees of freedom, in general,
introduces degenerate directions in phase space from the point of view of
symplectic geometry. Accordingly, the truncated phase-space geometry is
Poisson, making use of a Poisson bracket with a non-invertible Poisson tensor,
as opposed to an invertible symplectic form. This method was first described
in \cite{EffAc} within the setting of Poisson geometry, but with the benefit
of hindsight it had several independent precursors
\cite{VariationalEffAc,GaussianDyn,EnvQuantumChaos,QHDTunneling} that can now
be recognized as low-order versions of Poisson spaces.

As an instructive example, consider the uncertainty relation
\begin{equation} \label{Uncert}
 (\Delta x)^2(\Delta p)^2-C_{xp}^2\geq U
\end{equation}
where $U=\hbar^2/4$ in quantum mechanics. The fluctuations and covariance on
the left-hand side are part of a parameterization of states by moments. The
constant $U$ on the right-hand side can be seen as a parameter that controls
quantum features: It is positive for quantum states but may be smaller if the
moments belong to a classical state, as encountered for instance in classical
dynamics if it is expressed in term of the Koopman wave function. There are
three moments of second order for a single classical degree of freedom, all
seen on the left-hand side of the uncertainty relation (\ref{Uncert}). If one
aims to find a quasiclassical or phase-space description of these and only
these degrees of freedom, one has to work with an odd-dimensional phase space
that cannot be symplectic but may well (and indeed does) have a Poisson bracket.

Since moments are defined as polynomial expressions of basic expectation
values, we may introduce a Poisson bracket by first defining
\begin{equation}\label{Poisson}
\{\langle\hat{O}_1\rangle_{\rho},\langle\hat{O}_2\rangle_{\rho}\}= 
  \frac{\langle[\hat{O}_1,\hat{O}_2]\rangle_{\rho}}{i\hbar}
\end{equation}
for expectation values of operators in some state pure or mixed $\rho$. This
operation defines an antisymmetric bracket on expectation values
$\langle\hat{O}\rangle_{\rho}$, interpreted as a function on state space for
any fixed $\hat{O}$. The bracket also satisfies the Jacobi identity because
the commutator on the right-hand side does so. We can extend it
to products of expectation values, as needed for moments, by imposing the
Leibniz rule.

If we include only moments up to some finite order in our quasiclassical
description, (\ref{Poisson}) defines a Poisson bracket that, in general, is
not symplectic. Nevertheless, dynamics on this phase space is well defined if
the system is characterized by a Hamilton operator $\hat{H}$.
The standard equation
\begin{equation}
 \frac{{\rm d}\langle\hat{O}\rangle_{\rho}}{{\rm d}t} =
 \frac{\langle[\hat{O},\hat{H}]\rangle_{\rho}}{i\hbar}
\end{equation}
is then equivalent to Hamilton's equation
\begin{equation}
\frac{{\rm d}\langle\hat{O}\rangle_{\rho}}{{\rm d}t} =
  \{\langle\hat{O}\rangle_{\rho},H_{\rm eff}\}
\end{equation}
on phase space with the effective Hamiltonian
\begin{equation}
  H_{\rm eff}(\rho)=\langle\hat{H}\rangle_{\rho}\,,
\end{equation}
again interpreted as a function on the space of states.
In general, this equation couples infinitely many independent expectation
values to one another.

A truncation is therefore required for manageable approximations which,
depending on the truncation order of moments, defines the corresponding
effective description.  For instance, we may use a semiclassical approximation
of order $N$ in $\hbar$ if we include only expectation values
$\langle\hat{q}^a\hat{p}^b\rangle$ for $a+b\leq 2N$, $a\geq 0$ and $b\geq 0$.
It is more convenient to reformulate this truncation as using basic
expectation values $q=\langle\hat{q}\rangle$ and $p=\langle\hat{p}\rangle$
together with central moments in a completely symmetric (or Weyl) ordering,
\begin{equation}
 \Delta(q^ap^b)= \langle (\hat{q}-q)^a(\hat{p}-p)^b\rangle_{\rm symm}\,.
\end{equation}
This ordering is defined by summing all $(a+b)!$ permutations of
$(\hat{q}-q)^a(\hat{p}-p)^b$ and dividing by $(a+b)!$. For $a=b=1$, this
implies the standard covariance
\begin{equation}
  C_{qp}=\Delta(qp)=\frac{1}{2}\langle\hat{q}\hat{p}+\hat{p}\hat{q}\rangle-
  qp
\end{equation}
while, for instance,
\begin{eqnarray}
  \Delta(q^2p)&=&\frac{1}{3}\langle(\hat{q}-q)^2(\hat{p}-p)+(\hat{q}-q)(\hat{p}-p)(\hat{q}-q)
  +(\hat{p}-p)(\hat{q}-q)^2\rangle\nonumber\\
  &=&
      \frac{1}{3}\langle\hat{q}^2\hat{p}+\hat{q}\hat{p}\hat{q}+\hat{p}\hat{q}^2\rangle
      -\langle\hat{q}^2\rangle p- \langle\hat{q}\hat{p}+\hat{p}\hat{q}\rangle q +2q^2p
\end{eqnarray}

In this paper, we further analyze the underlying Poisson geometry and present
new results that demonstrate how well-known quantum parameters, such as $U$ in
(\ref{Uncert}), are equipped with an elegant geometrical interpretation as
centrifugal barriers for quantum degrees of freedom. Physical applications
related to tunneling will also be described, and we will compare these
methods with adiabatic or low-energy ones.

\section{Geometry of the free particle}

The quantum dynamics of the free particle can be described (in this case,
exactly) by the following geometrical picture. We have the effective
Hamiltonian
\begin{equation} \label{Hmoments}
H_{\rm eff}(p,\Delta(p^2))=
\left\langle\frac{\hat{p}^2}{2m}\right\rangle=
\frac{p^2}{2m}+\frac{\Delta(p^2)}{2m} \,.
\end{equation}
An immediate question in a quasiclassical picture is whether the last term,
$\Delta(p^2)/2m$ in which $\Delta(p^2)$ is often suggestively written as the
square $(\Delta p)^2$ of a momentum fluctuation $\Delta p$, is another kinetic
energy in addition to $p^2/(2m)$, or maybe contributes to the potential.

\subsection{Moment systems}

To answer this question, we consider the relevant
Poisson brackets
\begin{eqnarray}
  \{\Delta(q^2),\Delta(p^2)\}&=& 4\Delta(qp)\label{q2p2}\\
 \{\Delta(q^2),\Delta(qp)\}&=& 2\Delta(q^2)\\
  \{\Delta(qp),\Delta(p^2)\}&=&2\Delta(p^2)
\end{eqnarray}
of moments of the same order as $\Delta(p^2)$ in the new term of the effective
Hamiltonian. The Poisson bracket is defined in (\ref{Poisson}), also making use
of the Leibniz rule. For instance, to derive the first bracket, (\ref{q2p2}), we use
\begin{equation}
  \{\langle\hat{q}^2\rangle,\langle\hat{p}^2\rangle\}=
  \frac{\langle[\hat{q}^2,\hat{p}^2]\rangle}{i\hbar}=
  2\langle\hat{q}\hat{p}+\hat{p}\hat{q}\rangle
\end{equation}
directly from (\ref{Poisson}). The remaining terms obtained from expanding
\begin{eqnarray}
  \{\Delta(q^2),\Delta(p^2)\}&=&
                                 \{\langle\hat{q}^2\rangle-q^2,\langle\hat{p}^2\rangle-p^2\}\nonumber\\
  &=&
  \{\langle\hat{q}^2\rangle,\langle\hat{p}^2\rangle\}-
  \{\langle\hat{q}^2\rangle,p^2\}- \{q^2,\langle\hat{p}^2\rangle\}+
  \{q^2,p^2\}
\end{eqnarray}
require the Leibniz rule:
\begin{eqnarray}
  \{\langle\hat{q}^2\rangle,p^2\}&=&
                                       2p\{\langle\hat{q}^2\rangle,\langle\hat{p}\rangle\}=4qp\\
  \{q^2,\langle\hat{p}^2\rangle\} &=&
                                      2q\{\langle\hat{q}\rangle,\langle\hat{p}^2\rangle\}=
                                      4qp\\
  \{q^2,p^2\} &=& 4qp\{\langle\hat{q}\rangle,\langle\hat{p}\rangle\}=4qp\,.
\end{eqnarray}
Combining all terms, we obtain (\ref{q2p2}).

At generic orders, central moments have several welcome but also some unwelcome
properties. They are always real, by definition, facilitate the definition of
a general semiclassical hierarchy $\Delta(q^ap^b)\propto \hbar^{(a+b)/2}$, and
are always Poisson orthogonal to the basic expectation values:
$\{q,\Delta(q^ap^b)\}=0=\{p,\Delta(q^ap^b)\}$. However, complications can
sometimes arise because they are coordinates on a phase space with boundaries,
given by uncertainty relations including those of higher orders, and because
their Poisson brackets are not canonical. While there is a closed-form
expression for the Poisson bracket of two moments of canonical operators
$\hat{q}$ and $\hat{p}$,
given by \cite{EffAc,HigherMoments}
\begin{eqnarray}
\{\Delta(q^ap^b),\Delta(q^cp^d)\}&\!\!=\!\!&
a d \Delta(q^{a - 1}p^b) \Delta(q^cp^{d - 1}) - b c \Delta(q^ap^{b -
  1})  \Delta(q^{c - 1}p^d)\nonumber\\
&&+\sum_{n}
\left(\frac{i\hbar}{2}\right)^{n-1}
K_{abcd}^{n}\, \Delta(q^{a+c-n}p^{b+d -n})
\end{eqnarray}
where $1\leq n \leq {\rm Min}(a + c, b + d, a + b, c + d)$ and
\begin{equation}
K_{abcd}^n := \sum_{m= 0}^{n} (-1)^m m!(n-m)!
\left(\!\!\begin{array}{c}
a\\m
\end{array}\!\!\right)
\left(\!\!\begin{array}{c}
b\\n-m
\end{array}\!\!\right)
\left(\!\!\begin{array}{c}
c\\n-m
\end{array}\!\!\right)
\left(\!\!\begin{array}{c}
d\\m
\end{array}\!\!\right)\,,
\end{equation}
it is rather tedious to evaluate. Moreover, the non-canonical nature can make
it difficult to determine algebraic or physical properties, as already seen in the
example of the free particle.

The non-canonical nature of these brackets implies that $\Delta p$ is not an
obvious canonical momentum. However, the Casimir--Darboux theorem
\cite{Arnold} states that every Poisson manifold (that is, any space equipped
with a Poisson bracket) has local coordinates $q_i$, $p_i$, $C_I$ such that
$\{q_i,p_j\}=\delta_{ij}$ and $\{q_i,C_I\}=0=\{p_i,C_I\}$. The $q_i$ and $p_j$
are then pairs of canonical configuration coordinates and momenta, while the
Casimir variables $C_I$ are conserved by any Hamiltonian. (If the manifold is
symplectic, all $C_I$ are zero.)

It turns out that the brackets of second-order moments have 
Casimir--Darboux variables $(s,p_s,C)$ where
\begin{equation} \label{sps}
 \Delta(q^2) = s^2\quad,\quad \Delta(qp) = sp_s\quad,\quad
 \Delta(p^2) = p_s^2+\frac{C}{s^2}\,.
\end{equation}
The canonical bracket $\{s,p_s\}=1$ and $\{s,C\}=0=\{p_s,C\}$, required for
Casimir--Darboux variables, can directly be confirmed after inverting the
mapping to obtain
\begin{equation}\label{spsinverse}
  s=\sqrt{\Delta(q^2)}\quad,\quad
  p_s=\frac{\Delta(qp)}{\sqrt{\Delta(q^2)}}\quad,\quad
  C=\Delta(q^2)\Delta(p^2)-\Delta(qp)^2\,.
\end{equation}
The last equation shows that
\begin{equation} \label{C}
 C=\Delta(q^2)\Delta(p^2)-\Delta(qp)^2\geq \hbar^2/4
\end{equation}
is bounded from below by Heisenberg's uncertainty relation (or the
Schr\"odinger--Robertson version). These properties have been found
independently in a variety of contexts, including quantum field theory
\cite{VariationalEffAc}, quantum chaos \cite{EnvQuantumChaos}, quantum
chemistry \cite{QHDTunneling} and models of quantum gravity \cite{CQC}.

The canonical formulation (\ref{sps}) or (\ref{spsinverse}) can be found if
one combines the dynamics implied by the Hamiltonian (\ref{Hmoments}) with the
fact that $C$ as defined in terms of moments by (\ref{C}) is conserved
by any Hamiltonian that depends only on basic expectation values and
second-order moments; see for instance \cite{QHDTunneling}. Hamiltonian dynamics gives
\begin{equation}
  \frac{{\rm d}\Delta(q^2)}{{\rm d}t} = \{\Delta(q^2),H_{\rm
    eff}(p,\Delta(p^2))\}= \frac{2}{m}\Delta(qp)
\end{equation}
using the Poisson bracket (\ref{q2p2}). If we define $s=\sqrt{\Delta(q^2)}$,
we obtain
\begin{equation}
  \dot{s}=\frac{1}{2s} \frac{{\rm d}\Delta(q^2)}{{\rm d}t}=
  \frac{\Delta(qp)}{ms}
\end{equation}
and therefore $\Delta(qp)/s$ appears as a momentum $p_s$ of $s$, as in
(\ref{sps}). The equation for $\Delta(p^2)$ can then be obtained by solving
$C=\Delta(q^2)\Delta(p^2)-\Delta(qp)^2$ for this moment.

In general, however, the canonical structure is more difficult to discern and
requires more systematic derivations, for instance when one considers higher
moments or more than one classical pair of degrees of freedom. It may still be
possible to derive the momenta of some moments using Hamiltonian dynamics, but
deriving a complete mapping of all independent degrees of freedom to canonical
form requires more care. While there is no completely systematic procedure,
Poisson geometry helps to organize the derivation
\cite{Bosonize,EffPotRealize}. Again in the example of second-order moments
(\ref{sps}), we are still free to choose one degree of freedom as the first
canonical variable, $s=\sqrt{\Delta(q^2)}$. Any Poisson bracket with $s$ then
takes the form
\begin{equation}
  \{s,f\}= \frac{\partial f}{\partial p_s}
\end{equation}
with the momentum $p_s$ to be determined, where $f$ is any function of the
moments. By evaluating this equation for sufficiently many functions $f$, or
just for the basic moments, it provides differential equations in the
independent variable $p_s$, with $s$ as a parameter, that can be solved to
reveal the $p_s$-dependence of moments.

For instance, choosing $f=\Delta(qp)$, we obtain
\begin{equation}
  \frac{\partial\Delta(qp)}{\partial p_s}=\{s,\Delta(qp)\}= \frac{1}{2s}\{\Delta(q^2),\Delta(qp)\}=
  \frac{\Delta(q^2)}{s}=s\,.
\end{equation}
The solution, $\Delta(qp)=sp_s+ c(s)$ with a free $s$-dependent function
$c(s)$ is consistent with (\ref{sps}). The free function can be removed by a
canonical transformation, replacing $p_s$ with $p_s+c(s)/s$. For
$\Delta(p^2)$, we obtain the differential equation
$\partial\Delta(p^2)/\partial p_s= 2\Delta(qp)/s=2p_s$. The solution is again
consistent with (\ref{sps}), where the free $s$-dependent function is here
fixed by using $C$ as a conserved quantity.

If there are more than three moments, the procedure has to be iterated. After
finding the first canonical pair analogous to $(s,p_s)$, the next step
consists in rewriting the remaining moments in terms of quantities that have
vanishing Poisson brackets with both $s$ and $p_s$, and are therefore
canonically independent. For second-order moments of a single classical degree
of freedom, this step merely consists in identifying $C$ as a conserved
quantity, but it can be more challenging in higher-dimensional phase
spaces. At this point, the procedure is no longer fully systematic and usually
requires special considerations of the moments system and its Poisson
brackets in order to proceed in a tractable manner.

\subsection{Free dynamics}

In Casimir--Darboux variables, the effective Hamiltonian (\ref{Hmoments}) of the free particle
takes the form
\begin{equation}
 H_{\rm eff}(p,s,p_s; C)=
 \frac{p^2}{2m}+\frac{p_s^2}{2m}+\frac{C}{2ms^2}
\end{equation}
with, as we see now, contributions to both the kinetic and potential
energies. It may seem counter-intuitive that the free particle is subject to a
potential, but if the fluctuation direction with coordinate $s$ is included in
the configuration space, uncertainty relations must imply some kind of
repulsive potential that prevents $s$ from reaching zero. The new term
$C/(2ms^2)$ where $C\geq \hbar^2/4\not=0$ is precisely of this form.

Nevertheless, we can rephrase the dynamics as manifestly free if we
further extend our configuration space. We can interpret the $1/s^2$ potential
as a centrifugal one in an auxiliary plane with coordinates $(X,Y)$, such that
$X=s \cos\phi$, $Y=s\sin\phi$ with a spurious angle $\phi$. This
transformation is canonical if we relate
\begin{equation}
  p_s=\frac{Xp_X+Yp_Y}{\sqrt{X^2+Y^2}} \quad,
  \quad p_{\phi}=Xp_Y-Yp_X\,.
\end{equation}
Inverting the usual derivation of the centrifugal potential by transforming
from Cartesian to polar coordinates, we obtain the effective Hamiltonian in
the form
\begin{eqnarray} \label{HeffXY}
  H_{\rm eff}&=& \frac{p^2}{2m}+\frac{p_s^2}{2m}+
                 \frac{p_{\phi}^2}{2ms^2}\nonumber\\
  &=& \frac{p^2}{2m}+ \frac{(Xp_X+Yp_Y)^2+
      (Xp_Y-Yp_X)^2}{2m(X^2+Y^2)}\nonumber\\
  &=&                 \frac{p^2}{2m}+\frac{p_X^2}{2m}+\frac{p_Y^2}{2m}\,.
\end{eqnarray}

\begin{figure}
  \begin{center}
    \includegraphics[width=10cm]{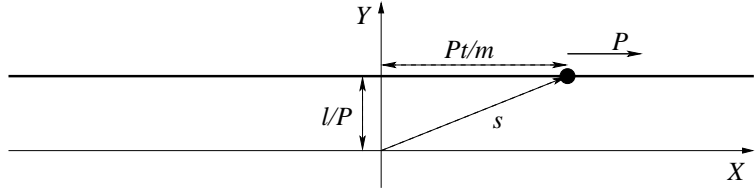}
    \caption{The auxiliary plane of free-particle motion. Since the effective
      Hamiltonian (\ref{HeffXY}) has no potential contribution, trajectories
      in the $(X,Y)$-plane are straight lines. Invoking rotational symmetry, a
      single trajectory may be chosen to point in the $X$-direction. Along
      this trajectory, $X(t)=Pt/m$ where $P$ is the conserved momentum in the
      $X$-direction, assuming the initial condition $X(0)=0$. The impact
      parameter of the trajectory relative to the origin equals angular
      momentum divided by the linear momentum, $l/P$. (The impact parameter is
      the distance $s(0)$ between the origin and the trajectory at a point
      where the trajectory is orthogonal to the radius vector. Therefore,
      $l=s(0)P$ or $s(0)=l/P$.) The right-angled triangle directly implies the
      solution (\ref{st}) for $s(t)$, also using $l=\sqrt{C}$ and the
      initial-value relationship $P=\sqrt{C}/s(0)$ that follows from the
      impact parameter.  \label{fig:Free}}
  \end{center}
\end{figure}

There is no potential, but trajectories still are not allowed to reach
$s=\sqrt{X^2+Y^2}=0$ because the interpretation of $C/(2ms^2)$ as a
centrifugal potential for motion in the $(X,Y)$-plane implies that the angular
momentum $l=p_{\phi}=\sqrt{C}\geq \hbar/2$ in the plane inherits a lower bound from
$C$. The uncertainty relation is therefore re-expressed as a centrifugal barrier
for motion that is required to have non-zero angular momentum in an auxiliary
plane. Going through the geometry of Fig.~\ref{fig:Free} shows that we obtain
the correct solutions
\begin{equation}\label{st}
 s(t)=s(0)\sqrt{1+\frac{Ct^2}{m^2s(0)^4}}
\end{equation}
depending on the constant $C\geq \hbar^2/2$, in agreement with quantum
fluctuations of a free particle. For a Gaussian state, $C=\hbar/2$ while
(\ref{st}) with $C>\hbar/2$ is also valid for non-Gaussian states.

\subsection{Two classical degrees of freedom}

For a pair of classical degrees of freedom, $x_1$ and $x_2$ with momenta $p_1$
and $p_2$, there are ten second-order moments. In terms of canonical
variables, the three position moments can be written as \cite{Bosonize,EffPotRealize}
\begin{equation}
 \Delta(x_1^2)=s_1^2 \quad,\quad \Delta (x_2^2)=s_2^2 \quad,\quad \Delta(x_1x_2)=
 s_1s_2\cos\beta
\end{equation}
with a new parameter $\beta$ that describes position correlations in the form of an angle.
The canonical momentum $p_{\beta}$ of $\beta$ appears in $\Delta(x_1p_2)$ and in
\begin{equation}
 \Delta(p_1^2)=p_{s_1}^2+ \frac{U_1}{s_1^2}
\end{equation}
where
\begin{equation}
 U_1=(p_{\alpha}-p_{\beta})^2+\frac{1}{2 \sin^2{\beta}}\left((C_1-4
  p_{\alpha}^2)-\sqrt{C_2-C_1^2+(C_1-4
    p_{\alpha}^2)^2}\sin{(\alpha+\beta)}\right)\,.
\end{equation}
Here, a fourth canonical parameter, $\alpha$, shows up together with its
momentum $p_{\alpha}$. The eight degrees of freedom
$(s_1,p_{s_1};s_2,p_{s_2};\alpha,p_{\alpha};\beta,p_{\beta})$ are completed to
ten independent degrees of freedom by two Casimir variables, $C_1$ and $C_2$
in $\Delta(p_1^2)$ and with a similar appearance in $\Delta(p_2^2)$.

The quasiclassical interpretation of uncertainty as rotation still holds for
two degrees of freedom: If we assume that $p_{\alpha}$ and $\sqrt[4]{C_2}$ are
much smaller than $p_{\beta}$ and $\sqrt{C_1}$, the dependence of $U_1$ on
$\alpha$ disappears, and we have
\begin{equation}
 \Delta (p_1^2)= p_{s_1}^2+ \frac{p_{\beta}^2}{s_1^2}+
 \frac{C_1}{2s_1^2\sin^2\beta} \,.
\end{equation}
This expression is equivalent to the kinetic energy in spherical coordinates
with angles $\vartheta=\beta$ and spurious $\varphi$, with constant amgular momentum
$p_{\varphi}=\sqrt{C_1/2}$.
Quantum uncertainty can therefore be modeled as a centrifugal barrier in a
3-dimensional auxiliary space with coordinates $(X,Y,Z)$ related to
$(s_1,\beta,\varphi)$ by a standard transformation between Cartesian and
spherical coordinates.

For this argument, we have to ignore the variables $\alpha$, $p_{\alpha}$ and
$C_2$. A few indications exist as to their possible physical meaning. First,
they turn out to be undetermined by a minimization of the effective potential
for two degrees of freedom subject to a generic classical potential
\cite{EffPotRealize}.  However, minimum energy should be realized in the
ground state, which should be given by unique wave function, a pure
state. Since moments refer to any state, pure or mixed, it is conceivable that
$\alpha$, $p_{\alpha}$ and $C_2$ describe moment degrees of freedom related to
the impurity of a state. To test this conjecture, one would have to work out
all relevant boundaries on the space of second-order moments, in addition to
the standard uncertainty relation for each classical pair. In order to produce
a unique and pure ground state, these boundaries would have to be such that the
ground-state moments are situated in a corner where $\alpha$, $p_{\alpha}$ and
$C_2$ can no longer vary when the energy is held fixed. These manifold
questions in a 10-dimensional phase space are quite tricky and remain to be
worked out. If the relationship with impurity can be made more precise, the
canonical variables would provide an interesting quasiclassical dynamics for
mixed states. The related effective potential could suggest new ways to
control impurity.

\section{Effective potentials}

In general, we may expand the effective Hamiltonian as
\begin{eqnarray}
H_{\rm eff}&=& \langle\hat{H}\rangle
= \langle
H(q+(\hat{q}-q),p+(\hat{p}-p))\rangle\\
&=& H(q,p)+ \sum_{a+b\geq 2} \frac{1}{a!b!} \frac{\partial^{a+b}H(q,p)}{\partial
  q^a\partial p^b} \Delta(q^ap^b) \nonumber
\end{eqnarray}
if $\hat{H}$ is given in completely symmetric ordering. (Otherwise, there will
be re-ordering terms that explicitly depend on $\hbar$.) 
The infinite series over $a$ and $b$ is reduced to a finite sum if $\hat{H}$ is polynomial in $\hat{q}$ and
$\hat{p}$. In this case, it just rewrites $\langle\hat{H}\rangle$ in terms of central
moments. For non-polynomial Hamiltonians, the series is expected to be
asymptotic rather than convergent.

If we truncate the moment order for semiclassical states, the series is also
reduced to a finite sum. This semiclassical interpretation is based on a broad
definition of semiclassical states where moments are assumed to be analytic in
$\hbar$, such that they obey the hierarchy
$\Delta(q^ap^b)\sim O(\hbar^{(a+b)/2})$. The interpretation does not
presuppose a specific shape of states, such as Gaussians.
If $\hat{H}=\frac{1}{2m}\hat{p}^2+V(\hat{q})$, the effective Hamiltonian
expanded to second order in moments reads
\begin{eqnarray}
 H_{\rm eff}(q,p,s,p_s)&=& \frac{p^2}{2m}+V(q)+
 \frac{1}{2m}\Delta(p^2)+\frac{1}{2}V''(q)\Delta(q^2)+ \cdots\nonumber\\
 &=& \frac{1}{2m}(p^2+p_s^2) +V(q)+ \frac{1}{2}V''(q)s^2
+ \frac{C}{2ms^2}+
 \cdots 
\end{eqnarray}
using as before Casimir--Darboux variables such that $\Delta(q^2)=s^2$ and
$\Delta(p^2)=p_s^2+C/s^2$.

\begin{figure}
  \begin{center}
    \includegraphics[width=14cm]{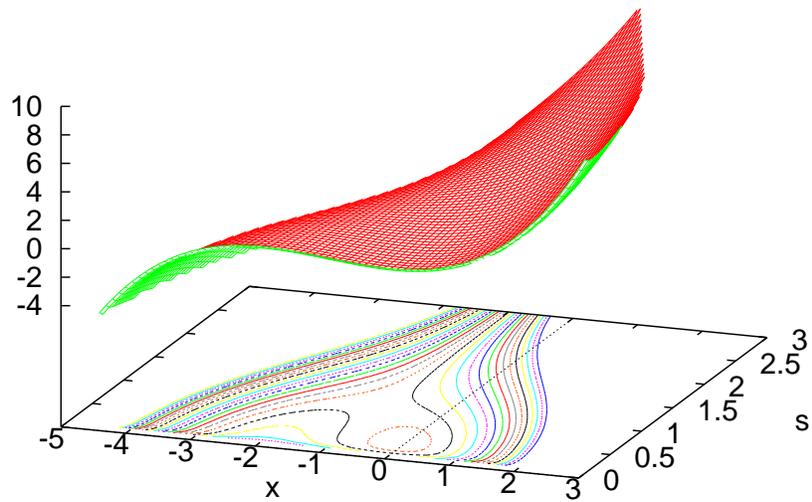}
    \caption{Second-order effective potential for a cubic classical potential. As the
      contour lines indicate, the classical barrier is lowered in the
      $s$-direction, making it possible for trajectories to bypass it while
      conserving energy. The lines also show that there is still a trapped
      region in this second-order truncation of the quantum
      potential. Higher-order moments would have to be included in order to
      describe complete tunneling at all energies.\label{fig:Cubics}}
  \end{center}
\end{figure}

An application to tunneling immediately follows because we always have
$V''(q)<0$ around local maxima of the potential, provided it is twice
differentiable. The term $\frac{1}{2}V''(q)s^2$ in the effective potential
therefore lowers the potential barrier in the new $s$-direction; see
Fig.~\ref{fig:Cubics} for an example.  Tunneling
can then be described by quasiclassical motion in an extended phase
space, bypassing the barrier with conserved energy
\cite{QHD,EffPotRealize}. (See also \cite{MomentsTunneling} for an analysis of
the same effect directly in terms of moments.)

This application of moment methods shows the importance of keeping $s$ as a
degree of freedom that evolves independently of $q$ (while being coupled to it
in a specific way). Other effective methods often replace independent degrees
of freedom such as $s$ with new $q$-dependent quantum corrections in an
effective Hamiltonian, which in general are of higher-derivative form. Such
higher-derivative or adiabatic corrections may be derived as a further
approximation within our quasiclassical systems. For instance,
the general equation
\begin{equation} \label{psdot}
  \dot{p}_s=-\frac{\partial H_{\rm eff}}{\partial s}
\end{equation}
provides a differential equation for $p_s$ depending on $q$ and $s$ on the
right-hand side. If we consider the full effective potential as in
Fig.~\ref{fig:Cubics}, the equation for $\dot{p}_s$ is part of the equations
of motion that describe a trajectory in the $(q,s)$-plane in which $s$ is kept
as a degree of freedom independent of $q$. Several other effective methods,
such as those based on path integrals as in \cite{EffAcQM}, do not exhibit new
independent degrees of freedom but rather use additional approximations that
(explicitly or implicitly) assume adiabatic evolution of quantum variables
such as $s$, while there is no such restriction on classical variables such as
$q$.

If the evolution of $s$ and its momentum $p_s$ is completely ignored, the
left-hand side of (\ref{psdot}) vanishes, and the equation is turned into an
algebraic equation relating $s=s_0(q)$ at an extremum of the effective
potential, where $\partial H_{\rm eff}/\partial s=0$. If this extremum is a
local minimum, the solution is stable, but it does not evolve in $s$ which
merely follows the evolution of $q$ in an adiabatic manner. If we insert
$s=s_0(q)$ as well as $p_s=m\dot{s}=m\dot{q} {\rm d}s_0(q)/{\rm d}q$ in the
effective Hamiltonian, we obtain a position-dependent mass correction of the
classical kinetic energy $\frac{1}{2}m\dot{q}^2$.

For small oscillations around the minimum, close to the ground state, one may
assume that $\dot{p}_s$ is small but not exactly zero, and include
corresponding deviations $\delta s$ of $s=s_0(q)+\delta s$ from its value at
the minimum. If one expands the right-hand side of (\ref{psdot}) in
$\delta s$, it implies a linear equation for $\delta s$ as a function of
$\dot{p}_s$, which in turn is related to $\ddot{s}$ by Hamilton's equation for
$s$. Inserting $s=s_0(q)+\delta s$ with the solutions for $s_0(q)$ and
$\delta s$ in the effective Hamiltonian implies a higher-derivative correction
in an effective Hamiltonian that now depends on second-order derivatives. At
higher orders in a systematic adiabatic expansion, derivatives of
arbitrary orders appear. In a second-order moment description, all these terms
are replaced by the coupled dynamics of a single quantum degree of freedom,
$s$.

\section{Conclusions}

We have presented several examples in which a canonical description of quantum
degrees of freedom can lead to new geometrical insights. We have presented
examples in which uncertainty relations are replaced by centrifugal barriers
in a quasiclassical phase space equipped with non-classical dimensions. Such
descriptions may make it easier to study the interplay between uncertainty
relations and the dynamics.

The novel formulation presented here, based on the mathematical subject of
Poisson geometry, unifies and extends several previous approaches from various
fields.  Quantum degrees of freedom, up to a given order in $\hbar$, are
represented by independent dynamical variables rather than higher-derivative
corrections of classical terms obtained from an adiabatic or derivative
expansion.  Physically, effective Hamiltonians that describe quantum evolution
by maintaining the classical number of degrees of freedom require higher
derivative terms because quantum dynamics is non-local in time (and also in
space in quantum field theory) if one considers only the classical degrees of
freedom. Because a wave function is in general spread out, it can have a
non-negligible effect on distant points even before one would expect the
classical position to reach there. By maintaining fluctuations and higher
moments as independent variables, however, the system can still be described
by local evolution.  The formalism described here therefore provides an
efficient and often intuitive description of what would appear as
non-adiabatic and non-local quantum effects in other effective descriptions.
If extensions to higher orders and multiple degrees of freedom are feasible,
there are promising advantages for numerical quantum evolution.

The methods also make it possible to characterize states and to distinguish
systematically between classical and quantum states, which may be of advantage
in hybrid treatments where some degrees of freedom can be considered
classical. Their moments can then be ignored, while other degrees of freedom
would couple to their own moments. As seen in our discussion of uncertainty
relations, Casimir variables also play a role because they are subject to
different bounds in classical and quantum systems. The standard uncertainty
relation has a lower bound of $C=0$ in classical physics because a
distribution on phase space may be sharply peaked in both $q$ and $p$, but
this is no longer possible in quantum physics. In this way, boundaries of
quasiclassical phase spaces provide a model independent way to distinguish
between classical and quantum systems or their hybrid combinations.

\section*{Acknowledgements}

The author thanks Cesare Tronci for an invitation to the 746.\
WE-Heraeus-Seminar ``Koopman Methods in Classical and Classical-Quantum
Mechanics'' where these results were presented.  This work was supported in
part by NSF grant PHY-2206591.


\end{document}